\documentclass[12pt]{article} 
\usepackage{amsfonts}
\usepackage{latexsym}
\usepackage{amsmath,amssymb}
\usepackage{verbatim}
\usepackage{mathtools}
\usepackage{color}
\usepackage{changebar}
\usepackage{setspace}
\usepackage{simplewick}
\usepackage[pdftex]{graphicx} 
\usepackage{tocvsec2}
\def\half{{1\over 2}}
\numberwithin{equation}{section}

 \def\p{\partial}

\newcommand{\bea}{\begin{eqnarray}}
\newcommand{\eea}{\end{eqnarray}}
\newcommand{\be}{\begin{equation}}
\newcommand{\ee}{\end{equation}}
\newcommand{\ba}{\begin{align}}
\newcommand{\ea}{\end{align}}

\newcommand{\W}{\mathcal{W}}

\newcommand{\hs}[1]{\mbox{hs$[#1]$}}
\newcommand{\w}[1]{\mbox{$\W_\infty[#1]$}}

\newcommand{\II}{\mathbb{I}}

\renewcommand{\P}[3]{{\cal P}_{#1}\!\left(\!\frac{#2}{#3}\!\right)}

\newcommand{\wP}{{\cal P}}
\renewcommand{\II}[3]{I_{#1}\left[\!\begin{smallmatrix} #2 \\ #3 \end{smallmatrix}\!\right]}

  \makeatletter
  \let\over=\@@over \let\overwithdelims=\@@overwithdelims
  \let\atop=\@@atop \let\atopwithdelims=\@@atopwithdelims
  \let\above=\@@above \let\abovewithdelims=\@@abovewithdelims

\DeclareMathOperator{\Tr}{Tr}

\begin{document}

\ \\

\begin{center}

{ \LARGE {\bf Higher Spin Black Holes from CFT }}

\vspace{0.8cm}

Matthias R.\ Gaberdiel$^{a}$, Thomas Hartman$^b$, and Kewang Jin$^a$
\vspace{1cm}

{\it $^a$Institut f\"ur Theoretische Physik, ETH Zurich, \\
CH-8093 Z\"urich, Switzerland \\
\tt{\small gaberdiel@itp.phys.ethz.ch}, \tt{\small jinke@itp.phys.ethz.ch}
 }

\vspace{0.5cm}

{\it $^b$Institute for Advanced Study, School of Natural Sciences,
\\
Princeton, NJ 08540, USA \\ 
\tt{\small hartman@ias.edu}
}

\vspace{1.0cm}

\end{center}

\begin{abstract}

Higher spin gravity in three dimensions has explicit black holes solutions, carrying higher spin charge.  We compute the free energy of a charged black hole from the holographic dual, a 2d CFT with extended conformal symmetry, and find exact agreement with the bulk thermodynamics.  
In the CFT, higher spin corrections to the free energy can be calculated at high temperature
from correlation functions of $\W$-algebra currents.

\end{abstract}

\pagestyle{empty}

\pagebreak
\setcounter{page}{1}
\pagestyle{plain}

\setcounter{tocdepth}{2}
\begin{singlespace}
\tableofcontents
\end{singlespace}

\section{Introduction}

Higher spin theories of gravity in Anti de Sitter space, with massless gauge fields of spin $s>2$ in addition to the graviton, are dual to field theories with extended conformal invariance \cite{Sundborg:2000wp,Mikhailov:2002bp,Sezgin:2002rt,Klebanov:2002ja,Giombi:2009wh,Maldacena:2011jn}. On the gravity side, higher spin gauge symmetries combine with diffeomorphisms, mixing the metric with the higher spin fields. Therefore a generalized notion of geometry is required in these theories, providing a simple toy model for string theory in the low tension regime where higher spin string states become massless. In the CFT, conserved higher spin currents dual to the bulk gauge fields impose powerful constraints that, in several examples, render the theory free or exactly solvable.

In AdS$_3$/CFT$_2$, the duality relates higher spin gravity to CFTs with $\W$-symmetry \cite{Henneaux:2010xg,Campoleoni:2010zq}. In particular, 3d Vasiliev gravity \cite{Fradkin:1986qy,Blencowe:1988gj,Vasiliev:1990en,Prokushkin:1998bq} is conjectured to be dual to the family of interacting but exactly solvable $\W_N$ minimal models \cite{Gaberdiel:2010pz}. In the large $N$ limit, the detailed microscopics of this proposal are supported by symmetries \cite{Gaberdiel:2011wb}, the spectrum of primaries \cite{Gaberdiel:2011zw}, and certain correlation functions
\cite{Chang:2011mz,Papadodimas:2011pf,Ahn:2011by,Chang:2011vk,Ammon:2011ua}; at finite $N$ some new ingredients may be required in the bulk (or perhaps can be found within the Vasiliev theory itself \cite{Castro:2011iw}).

In this paper we focus on a universal aspect of the correspondence: black holes in AdS$_3$ carrying higher spin charge. These black holes encode the thermodynamics of the CFT at high temperature, which in two dimensions is determined by the chiral algebra. Thus black holes are insensitive to some of the microscopic details of the CFT, but capture certain features of the symmetries, as well as the asymptotic density of states.

It is not obvious how to define a black hole in higher spin gravity. The usual definition --- a spacetime singularity hidden behind a horizon --- is difficult to apply because neither the Riemann tensor nor the causal structure of the metric are gauge invariant. In Euclidean signature, the problem is simpler, because a black hole is simply a smooth classical solution with torus boundary conditions. This definition has been used to construct explicit black hole solutions carrying higher spin charge \cite{Gutperle:2011kf} (see also \cite{Ammon:2011nk,Castro:2011fm,Tan:2011tj}). An interesting feature of these constructions is that turning on a chemical potential for a higher spin charge means deforming the CFT by an irrelevant operator, so these black holes describe an RG flow to a UV fixed point. The flow can be solved exactly in the bulk, resulting in a new AdS geometry in the UV that is realized by a different subset of the bulk gauge fields.

There are many different theories of higher spin gravity in AdS$_3$, specified by a Lie algebra.  The original black holes of \cite{Gutperle:2011kf} are solutions of $sl(N)$ gravity, which has gauge fields up to spin $N$.  We will consider higher spin gravity based on the infinite-dimensional Lie algebra $\hs{\lambda}$, since this is the bulk theory in the minimal model duality.  (It is also in a sense universal, because setting $\lambda=N$ gives $sl(N)$ gravity.)  Black hole solutions in this theory were constructed in \cite{Kraus:2011ds}.  The mass, angular momentum, and charges of the black hole were computed and used to infer the free energy \cite{Kraus:2011ds}
\be\label{gravres}
\log Z_{\rm BH}(\hat\tau, \alpha) = \frac{i\pi c}{12\, \hat\tau} \Bigl[ 
1 - \frac{4}{3} \frac{\alpha^2}{\hat\tau^4} + \frac{400}{27} \frac{\lambda^2-7}{\lambda^2-4} \, 
\frac{\alpha^4}{\hat\tau^8} -\frac{1600}{27}\frac{5\lambda^4-85\lambda^2+377}{(\lambda^2-4)^2}\frac{\alpha^6}{\hat{\tau}^{12}}+ \cdots \Bigr] \ ,
\ee
where $\alpha$ is the chemical potential for spin-3 charge, $\hat{\tau}$ is related to the mass and angular momentum (see below), and $c=3\ell/2G$ with $\ell$ the AdS radius and $G$ the Newton's constant.\footnote{We display only the holomorphic piece of the partition function; the right-moving sector gives a similar contribution, in gravity and in CFT.} The free parameter $\lambda$ is a choice of theory, labelling the higher spin algebra $\hs{\lambda}$. Our aim is to reproduce this formula from CFT; the usual thermodynamic relation also gives the entropy, $S = (1-\beta \p_\beta)\log Z$. In \cite{Kraus:2011ds}, the formula (\ref{gravres}) was already derived from CFT at the special values $\lambda=0,1$, where the chiral algebra has a simple free field realization, but for arbitrary $\lambda$ a more general approach is needed.

Black hole entropy was first derived from microstate counting in string theory by Strominger and Vafa, using the detailed microscopic description of D-branes \cite{Strominger:1996sh}.  Later, this computation was streamlined and generalized by showing that the asymptotic density of D-brane microstates can be counted indirectly using Cardy's formula, for which the only input is the central charge of the CFT \cite{Strominger:1997eq}.  We will derive a similar universal formula for the thermodynamics of higher spin CFTs, and show that it exactly matches the gravity result (\ref{gravres}).

The general strategy is as follows.  We start with the partition function of the CFT,
\be
Z_{\rm CFT} = \Tr \Bigl( e^{-\beta H + 2\pi i \alpha W_0} \Bigr) \ ,
\ee
where $W_0$ is the zero mode of the spin-3 current.  The CFT has $\w{\lambda}$ symmetry, an extension of the Virasoro algebra by an infinite number of higher spin currents.  The partition function is expanded perturbatively in $\alpha$; then a modular transformation is applied to each term, which makes it possible to compute the trace at high temperature.  Thus the higher-spin corrections to the partition function are accounted for by certain correlation functions of $W$'s on a torus. Unlike the bulk calculation, where only $\hs{\lambda}$ is used explicitly, the CFT result depends in an essential way on the nonlinear terms in \w{\lambda}. 

The computation relies only on the chiral algebra of the CFT, and does not require (or allow for) explicit enumeration of the actual microstates contributing to the entropy.  This means that although the minimal model CFTs appearing in the duality proposed in \cite{Gaberdiel:2010pz} certainly reproduce the black hole entropy at high enough temperatures, it is not a fine-grained test of the specific details of that CFT; we will return to this point in the discussion section. 

In section \ref{sec:strategy}, we explain in detail the strategy of the computation, and state our main technical tool, a recursion relation for the evaluation of torus correlation functions.  In section \ref{s:apply}, we apply the recursion relation and compare to the gravity answer (\ref{gravres}).  In section \ref{s:discussion}, we conclude with a discussion of the regime of validity of the formula and applicability to minimal model holography.  Various technical details are relegated to the appendices: appendix \ref{sec:Weierstrass} defines the Weierstrass functions that appear in torus amplitudes; in appendix \ref{sec:recursion}, we define some notation and derive the recursion relation for torus amplitudes;  details of the six-point calculation are given in appendix \ref{a:sixpoint}; and the commutation relations of $\w{\lambda}$ appear in appendix \ref{a:commutators}.

\section{Higher Spin Corrections to CFT Thermodynamics}\label{sec:strategy}

Let us begin by explaining the aim of our calculation in more detail. The entropy calculation of
\cite{Kraus:2011ds} amounts, from the dual CFT point of view, to the evaluation of the
partition function
\be\label{ZAdS}
Z_{\rm CFT}(\hat\tau, \alpha) = \Tr \Bigl( \hat{q}^{L_0 - \frac{c}{24}} \, y^{W_0} \Bigr)  \ , \qquad
\hat{q} = e^{2\pi i \hat{\tau}} \ , \quad y = e^{2\pi i \alpha} \ ,
\ee
where the trace is taken over the entire spectrum.
Here $W_0$ is the zero mode of the spin $3$ generator of $\w{\lambda}$, with chemical potential $\alpha$, while
$\hat\tau$ is the complex structure of the torus, related to the black hole temperature $T_H$ and (imaginary) angular potential $\Omega_H$ by
\be
\hat{\tau} = \frac{i}{2\pi T_H}(1 + \Omega_H) \ .
\ee

The bulk partition function (\ref{gravres}) was obtained in \cite{Kraus:2011ds} from the thermodynamics of the black hole solution in AdS$_3$ higher spin gravity.  It is an expansion in powers of the chemical potential $\alpha$, so it should be compared to the CFT expansion
\be\label{gravex}
Z_{\rm CFT}(\hat\tau, \alpha) = \Tr \Bigl( \hat{q}^{L_0 - \frac{c}{24}} \Bigr) + 
\frac{(2\pi i \alpha)^2}{2!}  \Tr \Bigl( (W_0)^2\, \hat{q}^{L_0 - \frac{c}{24}} \Bigr) + 
\frac{(2\pi i \alpha)^4}{4!}  \Tr \Bigl( (W_0)^4\, \hat{q}^{L_0 - \frac{c}{24}} \Bigr) + \cdots 
\ee
in the high temperature regime, i.e.\ for $\hat\tau\rightarrow 0$, and to leading order in the central charge $c$ (since the gravity calculation is only reliable for large $c$). At high temperatures, the $\hat{\tau}$-dependence of each term in the expansion is fixed by conformal invariance, which requires\footnote{Reintroducing the size of the 
system $R$, the dimensionful temperature $T\sim i /(\hat{\tau} R)$ and a chemical potential 
$\bar{\alpha} \sim \alpha R^2$ of dimension $-2$, as appropriate for a spin-3 charge, 
the only dimensionless quantity is ${\alpha}/{\hat\tau^2}$. Since 
$\exp(2\pi i \alpha W_0)$ defines, at least formally, a unitary operator, the asymptotic 
behaviour of $Z_{\rm CFT}(\hat{\tau}, \alpha)$ must be the same as for the usual trace,
 i.e.\ proportional to $\hat\tau^{-1}$, and thus the structure must be as in 
(\ref{taudep}). Alternatively, we may argue that $\log Z$ can be at most linear in 
the system size $R$ since the free energy per unit size should be finite.}
\be\label{taudep}
\log Z_{\rm CFT}(\hat{\tau}, \alpha) \approx \frac{1}{\hat{\tau}}f\left(\alpha\over \hat{\tau}^2\right)
\ee
for some function $f$. 

As is familiar from entropy calculations \cite{Strominger:1997eq}, the standard method to obtain the partition function from
a dual conformal field theory point of view is to do the $S$-modular transformation
\be
\tau = -\frac{1}{\hat{\tau}} \ , \qquad q = e^{2\pi i \tau} \ .
\ee
In the high temperature regime, i.e.\ 
for $\hat\tau\rightarrow 0$,  we have $q\rightarrow 0$. The answer for the trace is then
dominated by the contribution from the vacuum state. 

This argument can be directly applied to the first term in the expansion 
(\ref{gravex}), 
\be
\Tr  \Bigl( \hat{q}^{L_0 - \frac{c}{24}} \Bigr) = \sum_{s,r} 
S_{s r}  \Tr_r \Bigl( q^{L_0-\frac{c}{24}} \Bigr) 
\sim \left(\sum_s S_{s 0}\right) \, q^{-\frac{c}{24}} + \cdots \ , 
\ee
where the sum runs over all primaries labelled by $r,s$ (with $r=0$ the 
vacuum representation), $S_{sr}$
is the modular $S$-matrix (not to be confused with the black hole entropy), and the dots indicate terms exponentially suppressed at high temperature. The
leading behaviour of the logarithm is then
\be
\log \Tr \Bigl( \hat{q}^{L_0 - \frac{c}{24}} \Bigr)  = - \frac{ i \pi c }{12} \tau + \cdots \ ,
\ee
and this reproduces precisely the $\alpha$-independent term in (\ref{gravres}),
using the relation $\tau=-\tfrac{1}{\hat\tau}$. This is equivalent to the Cardy formula for the entropy.  In the following we want to repeat this analysis for the other terms in the expansion
(\ref{gravex}). The main technical problem we need to understand is how traces
with zero mode insertions behave under the modular $S$-transformation. 
\smallskip

Consider first the simpler case where the zero mode is that of a dimension-1 current, and the modular transformation
of the full exponential is actually known to be described by a Jacobi form.  (This results in a simple Cardy-like formula for the entropy of a black hole carrying $U(1)$ charge.)  Let us assume
that we have a rational CFT with a $U(1)$ current $J_n$, and define the
trace in the representation labelled by $r$ 
\begin{equation}
\phi_r(\tau,z)={\rm Tr}_r(y^{J_0} q^{L_0-{c \over 24}})\ , \qquad y=e^{2\pi i z}\ , 
\quad q=e^{2\pi i \tau} \ .
\end{equation}
Then it follows from usual CFT arguments\footnote{In particular, this argument relies on the 
spectral flow automorphism of the $U(1)$ current algebra, see for example \cite{Kraus:2006wn}. } that  
\begin{equation}\label{jacobi}
\phi_r \left({a\tau +b \over c\tau +d},{z \over c\tau +d} \right)= \sum_s
\exp\Bigl\{2\pi i k {c z^2 \over c\tau +d}\Bigr\} \, \,
M_{rs}\left( \begin{matrix} a & b \\ c & d \end{matrix} \right) \,
\,  \phi_s(\tau,z)\ , 
\end{equation}
where the sum runs over all irreducible representations of the chiral algebra, and
$M_{rs}$ defines a representation of the modular group. Furthermore, the level $k$
is defined via the commutator
\begin{equation}
{}[J_m,J_n] = 2 k\, m \delta_{m,-n} \ .
\end{equation}
Expanding in powers of $z$, one then finds for example that 
under the $S$-modular transformation
\be
{\rm Tr}_r \Bigl( J_0 J_0 \hat{q}^{L_0-\frac{c}{24}} \Bigr) = 
\sum_s S_{rs} \Bigl[ \tau^2\, 
{\rm Tr}_s \Bigl( J_0 J_0 q^{L_0-\frac{c}{24}} \Bigr)  + 
\frac{k}{\pi i} \,\tau\, {\rm Tr}_s \Bigl( q^{L_0-\frac{c}{24}} \Bigr) \Bigr]  \ .
\ee
In the high temperature $q\rightarrow 0$ limit, it is 
the second term evaluated for $s=0$ that is dominant. The zero modes inside the first trace
annihilate the vacuum so it is exponentially suppressed.

Unfortunately, the analogue of (\ref{jacobi}) for zero modes of general spin
is not known. There have been some ideas that the relevant transformation
property will be of the same structure as that described in \cite{KZ}, see
\cite{Wang}, but no general formula is known.\footnote{We thank Jan Manschot for
bringing these papers to our attention.} As we shall explain
in the following, we can derive the transformation properties of these
traces from first principles.

\subsection{The general strategy}

In order to explain the general strategy let us introduce a little bit 
of notation. For each representation of the chiral algebra (labelled by $r$), we can define
a torus amplitude by 
\begin{equation}\label{Fdef}
F_r((a^1,z_1),\ldots,(a^n,z_n);\tau)=z_1^{h_1} \cdots z_n^{h_n} \, {\rm Tr}_r
\Bigl(V(a^1,z_1) \cdots V(a^n,z_n)\, q^{L_0-{c \over 24}}\Bigr) \ ,
\end{equation}
where $h_j$ are the conformal dimensions of the chiral fields $a^j$, i.e.\ $L_0 a^j = h_j a^j$
with $h_j\in {\mathbb N}$. 
(In our case, the chiral fields will be the higher spin fields of the $\w{\lambda}$ algebra.)
These functions are periodic in each variable $z_j$ under the transformations
\be\label{peri}
z_j \mapsto e^{2\pi i } z_j \ , \qquad \hbox{and} \qquad z_j \mapsto q z_j  \ ,
\ee
and therefore deserve to be called `torus amplitudes'. 
The periodicity under $z_j \mapsto e^{2\pi i } z_j$ is obvious; to see the
periodicity under $z_j \mapsto q z_j$ we first note that the functions are independent
of the order in which the vertex operators are inserted (since they are local fields). 
Using the transformation property of the fields under scaling
\be
V(a^j,q\, z_j) \, q^{L_0-\frac{c}{24}}  = q^{-h_j}\, q^{L_0-\frac{c}{24}}  \, V(a^j,z_j) \ ,
\ee
as well as the cyclicity of the trace it follows that the functions in (\ref{Fdef}) are also
invariant under $z_j \mapsto q z_j$.

The amplitudes have a simple transformation property under the modular group provided
that all $a^j$ are Virasoro primaries so that $L_n a^j =0$ for $n>0$. If this is the case then under a modular transformation \cite{Zhu}
\begin{equation}\label{Ftrans}
F_r((a^1,z_1),\ldots,(a^n,z_n);\tfrac{a\tau + b }{c \tau + d}) = (c\tau +d)^{\sum_j h_j} \, 
\sum_s M_{rs}  \,
F_s((a^1,z_1^{c\tau+d}),\ldots,(a^n,z_n^{c\tau+d});\tau) \ ,
\end{equation}
where $M_{rs}$ is again a matrix representation of the modular group, i.e.\ it depends on the 
matrix $\left(\begin{smallmatrix} a & b \\ c & d \end{smallmatrix} \right)$, but not on the $a^i$. 
Note that this transformation rule makes sense: the left-hand-side is invariant under 
\be
z_j \mapsto e^{2\pi i } z_j \ , \qquad \hbox{and} \qquad z_j \mapsto \tilde{q} z_j  \quad
(\hbox{with} \quad \tilde{q} = e^{2\pi i \frac{a\tau +b}{c \tau +d}} )\ , 
\ee
under which $z_j^{c\tau + d}$ transforms as 
\be
z_j^{c\tau +d} \mapsto e^{2\pi i (c\tau +d)} z_j^{c\tau +d} \ , \qquad 
\hbox{and} \qquad z_j^{c\tau+d} \mapsto e^{2\pi i (a\tau + b)} z_j^{c\tau +d} \ ,
\ee
respectively. These are both periodicities of the right-hand-side, see (\ref{peri}).
\smallskip

\begin{figure}[t]
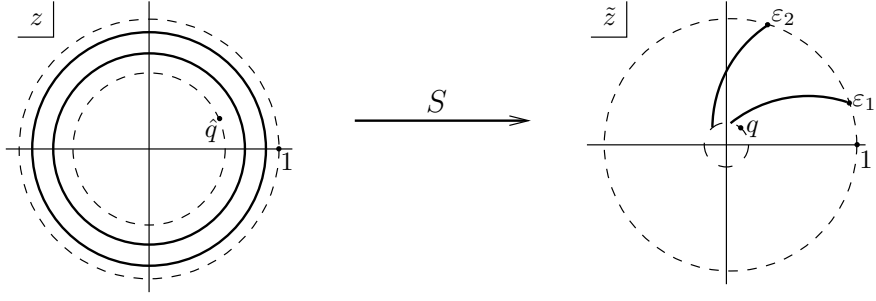

\begin{center}\input contours2.pdftex_t\end{center}
\hspace{.5in}\parbox{5.5in}{\caption{\label{fig:contours}{ \small Contours of integration for the insertion of two zero modes, before and after the $S$ modular transformation.  The fundamental region of the torus lies between the dashed lines, $|q| < |z| \leq 1$.} }}
\end{figure}

We are interested in the modular transformation properties of the traces with the insertion of 
zero modes. Expanding the vertex operators in modes as 
\begin{equation}
V(a,z)=\sum_{m \in \mathbb{Z}} a_m \, z^{-m-h}\ ,
\end{equation}
where $h$ is the conformal dimension of $a$, 
the zero modes are obtained from the torus amplitude via the integrals $\oint \frac{dz}{z}$, i.e.\
\be\label{zero}
\Tr_r \Bigl( a^1_0 \cdots a^n_0 \, q^{L_0-\frac{c}{24}} \Bigr) = 
\frac{1}{(2\pi i)^n} \oint \frac{dz_1}{z_1} \cdots \oint \frac{dz_n}{z_n}\, F_r((a^1,z_1),\ldots,(a^n,z_n);\tau)  \ .
\ee
Since we know the modular transformation properties of the torus amplitudes, this can 
now be used to deduce that of the traces with zero modes. Concentrating on the $S$-modular
transformation that is of relevance for us, we thus obtain from (\ref{zero}) and (\ref{Ftrans})
\begin{equation}
(2\pi i)^n\Tr_r\Bigl(a^1_0 \cdots a^n_0 \, \hat{q}^{L_0-\frac{c}{24}} \Bigr) =
 \sum_{s} S_{rs}\, \tau^{-n+\sum_j h_j} \!\!
 \int_1^q \frac{d\tilde{z}_1}{\tilde{z}_1} \cdots \!\! \int_1^q \frac{d\tilde{z}_n}{\tilde{z}_n}\,
F_s((a^1,\tilde{z}_1),\ldots,(a^n,\tilde{z}_n);\tau) \ . \label{form1} 
\end{equation}
Here the integration contours are the result of the transformation $z\rightarrow z^\tau\equiv \tilde{z}$, which swaps the two cycles of the torus under the $S$-transformation, see figure \ref{fig:contours}. The contours are displaced to avoid short-distance singularities, so the contour for $\tilde{z}_j$  runs from $\varepsilon_j$ to $\varepsilon_j q$ for some arbitrary phase $\varepsilon_j$; these phases are implicit throughout. From now on we work on the $\tilde{z}$-plane and drop the tildes.

In the high temperature limit  $\hat\tau\rightarrow 0$ the dominant contribution 
to (\ref{form1}) will come from the vacuum $s=0$. 
Thus our calculation is reduced to doing the multiple integrals of the torus amplitudes in 
(\ref{form1}) with $s=0$. In order to evaluate the torus amplitudes we can use the 
recursion relation of \cite{Zhu} to rewrite (from now on we drop the index $s=0$ from
the torus amplitudes and the traces)
\begin{eqnarray}
& F & \hspace*{-0.2cm} 
\Bigl( (a^1,z_1),(a^2,z_2),\ldots,(a^n,z_n);\tau \Bigr)
 \nonumber \\
& = & \hspace*{-0.2cm}
z_2^{h_2}\cdots z_n^{h_n} 
\Tr \Bigl( a^1_0\, V(a^2,z_2)\cdots  V(a^n,z_n)\, q^{L_0-\frac{c}{24}} \Bigr) \label{recur}  \\
& & \hspace*{-0.2cm}
+ \sum_{j=2}^{n}\, \sum_{m\in\mathbb{N}_0} \, 
{\cal P}_{m+1}\left(\frac{z_j}{z_1},q\right) \, 
F\Bigl( (a^2,z_2),\ldots,
(a^1[m]a^j,z_j),\ldots , (a^n,z_n);\tau \Bigr) \ ,\nonumber
\end{eqnarray}
where ${\cal P}_{m+1}(x,q)$ are the Weierstrass functions defined in 
appendix~\ref{sec:Weierstrass}, and 
the bracketed modes, defined in appendix~\ref{sec:brecketed}, have the form
\be
a[m] = (2\pi i)^{-m-1}\sum_{i=m}^\infty c(h_a,i,m)a_{-h_a + 1 + i} \ .
\ee
Note that both sides are periodic under $z_i \mapsto  q \,z_i$;
this is manifest for all terms except for the term in the second line, and the 
$m=0$ terms in the double sum of the last line, given that ${\cal P}_1$ transforms as 
(see eq.~(\ref{Pperiod}))
\be\label{ano}
{\cal P}_{1}\left(\frac{q z_i}{z_1},q\right) 
= {\cal P}_{1}\left(\frac{z_i}{z_1},q\right) + 2 \pi i \ . 
\ee
It is not difficult to check that the non-periodic contribution of  (\ref{ano}) cancels precisely
the non-trivial term that appears from the second line, where we pick up a commutator term
from moving $V(a^i,{z}_i)$ past $a^1_0$. Indeed, this commutator is precisely equal to 
\be\label{zeroco}
[a^1_0, V(z^{L_0} \, a^i,z) ] =  (2\pi i)\, V(z^{L_0}\, a^1[0]a^i, z) \ ,
\ee
so that
\be\label{fcom}
F\left(\dots,[a_0, V(a^i,z)],\dots\right) = (2\pi i)\,F\left(\dots,V(a[0]a^i,z),\dots\right) \ .
\ee

Unfortunately, this simple recursion relation is not quite sufficient (for $n>2$) since
the term in the second line involves an explicit zero mode, and therefore cannot
be recursed further in this manner. We have therefore generalised the analysis of
\cite{Zhu} to find also recursion relations for correlators involving zero modes. Defining
\be
F\left(b_0^\ell; (a^1,z_1),\dots,(a^n,z_n); \tau\right) 
= z_1^{h_1}\cdots z_n^{h_n}\,
\Tr\Bigl( b_0^\ell \, V(a^1,z_1)\dots V(a^n,z_n)\, q^{L_0-\frac{c}{24}} \Bigr)
\ee
we have
\begin{align}\label{multrec}
F\big(b_0^\ell;& (a^1,{z}_1),\dots,(a^n,{z}_n);\tau\big) =
F\left(b_0^\ell a^1_0; (a^2,{z}_2),\dots,(a^n,{z}_n);\tau\right)\\
& + \sum_{i=0}^{\ell} \sum_{j=2}^n \sum_{m\in\mathbb{N}_0}  
{\ell \choose i}\, g_{m+1}^i\!\left({z}_j\over {z}_1\!\right)
F\left(b_0^{\ell-i}; (a^2,{z}_2),\dots,(d^{(i)}[m]a^j,{z}_j),\dots,(a^n,{z}_n);\tau\right) \ ,
\notag
\end{align}
where
\be\label{gderiv}
g^i_{m+1}(x,q) = (2\pi i)^i \frac{(m-i)!}{m!}\p_\tau^i {\cal P}_{m+1-i}(x,q) \qquad (m\geq i) 
\ee
and
\be
d^{(i)} = (-1)^i (b[0])^i a^1 \ .
\ee
This identity is proven in appendix~\ref{sec:recursion}.\footnote{ The function $g_{m+1}^i(x,q)$ is extended to the range $m<i$ in appendix \ref{sec:Weierstrass}. This is necessary for the general recursion relation if zero modes of the chiral algebra do not commute, but will not appear in any of our computations.} Note that the sum over $m$ is actually finite because modes with high enough 
mode number will annihilate $a^j$.

With the help of this recursion relation we can then evaluate the amplitudes in (\ref{form1})
explicitly (and recursively).  The recursion formula turns the amplitude into a sum of products of Weierstrass functions; then the contour integrals are straightforward because ${\cal P}(z)/z$ is a total derivative. This computation will be explained in detail for the $(W_0)^2$, $(W_0)^4$, and $(W_0)^6$ cases in the following section. Fortunately, we do not need to evaluate the exact answer, but only the leading 
order contribution as $c\rightarrow \infty$.

\section{Applying the Recursion Relation}\label{s:apply}

Exponentiating the gravity result (\ref{gravres}) leads to 
\be\label{gravcon}
Z_{\rm BH} = q^{-c/24}\left[1  + 2\pi i c\tau  \left(\frac{1}{18}\alpha^2 \tau^4 - \frac{50}{81}\frac{\lambda^2-7}{\lambda^2-4}\alpha^4\tau^8 + \frac{200}{81}\frac{5\lambda^4 - 85\lambda^2 + 377}{(\lambda^2-4)^2}\alpha^6\tau^{12}\right)+ \cdots\right] \ ,
\ee
where we have written only the order $c$ part of the expansion. Higher powers in $c$ are given by disconnected diagrams, so we will compare (\ref{gravcon}) to the connected contribution to the CFT partition function.  To compute connected correlators, we simply discard terms where some subset of the $W$'s produce a central term at an intermediate step in the calculation. In this section we will use the recursion relation to compute correlation functions of $W$ currents and thereby reproduce (\ref{gravcon}) in CFT, term by term.

\subsection{The 2-point case}

The first higher-spin correction to the partition function is independent of $\lambda$, and was derived from CFT in \cite{Kraus:2011ds} by taking advantage of the free-field realization of $\w{\lambda}$ at $\lambda=0,1$. We will rederive it by the strategy outlined above, in order to illustrate the method.

The leading correction in the expansion (\ref{gravex}) is, using (\ref{form1}), 
\be
Z^{(2)} \equiv \frac{(2\pi i\alpha)^2}{2!}\Tr\bigl( (W_0)^2 \hat{q}^{L_0-\frac{c}{24}} \bigr)
\approx \frac{\alpha^2\tau^4}{2}\,
 \int_1^q\frac{dz_1}{z_1}\int_1^q\frac{dz_2}{z_2}F\big((W,z_1), (W,z_2); \tau\big) \ .
\ee
The symbol `$\approx$' means we have dropped terms that do not contribute to $\log Z$ at leading order. Applying the recursion relation (\ref{recur}), 
\be
F\big((W,z_1), (W,z_2); \tau\big) = z_2^3 \Tr\bigl( W_0 W(z_2)\, q^{L_0-\frac{c}{24}} \bigr)
+ {\cal P}_{m+1}\left(z_2\over z_1\right) F\big((W[m]W, z_2); \tau\big) \ .
\ee
Here, as in the following, we shall always imply a sum
over the Weierstrass index $m \geq 0$. In the first term, only the zero mode of $W(z_2)$ 
survives in the trace, but $\Tr \bigl((W_0)^2 q^{L_0-c/24}\bigr)$ has no contribution from the 
vacuum state and hence is exponentially suppressed.  In the second term, the vacuum 
contribution to the trace dominates, so 
\be
Z^{(2)} \approx \half q^{-c/24} \alpha^2 \tau^4 \langle W[m]W_{-3} \rangle 
\, \int_1^q\frac{dz_1}{z_1}\int_1^q\frac{dz_2}{z_2} {\cal P}_{m+1}\left(z_2\over z_1\right) \ ,
\ee
where $\langle \cdot \rangle$ is the vacuum expectation value (on the sphere).  The term with $m=0$ vanishes by (\ref{fcom}).  According to (\ref{Prec}), integrating a Weierstrass function reduces its index by 1. Only ${\cal P}_1$ is not periodic, so this implies
\be
\int_1^q \frac{dz_2}{z_2}\, {\cal P}_2\left(z_1\over z_2\right) = (2\pi i)^2 \  , \quad 
\int_1^q \frac{dz_2}{z_2}\, {\cal P}_{m+1}\left(z_1\over z_2\right) = 0 \quad (m>1) \ .
\ee
Therefore
\be
Z^{(2)}\approx \half q^{-c/24}(2\pi i)^3 \alpha^2 \tau^5 \langle W[1]W_{-3}\rangle \ .
\ee
The definition of the bracketed mode (\ref{abradef}) is
\be
W[1] = (2\pi i)^{-2}\left(W_{-1} + \frac{3}{2}W_0 + \frac{1}{3}W_{1} - \frac{1}{12}W_{2} + \frac{1}{30}W_{3} + \cdots \right) \ .
\ee
Since we are only interested in the vacuum expectation value, the only surviving term is then
$\langle W[1]W_{-3}\rangle = (2\pi i)^{-2}\tfrac{1}{30} \langle W_3 W_{-3} \rangle$, and plugging in 
the commutators of $\w{\lambda}$ from appendix \ref{a:commutators}, the final answer is 
\be
Z^{(2)} = 2 \pi i c\, \frac{\alpha^2\tau^5}{18} \,q^{-c/24}\ ,
\ee
in precise agreement with the leading correction to the gravitational partition 
function (\ref{gravcon}). 

\subsection{The 4-point function}\label{ss:fourpoint}

Now we proceed to the $\alpha^4$ correction to the partition function,
\be\label{z4}
Z^{(4)} \equiv \frac{(2\pi i \alpha)^4}{4!}
\Tr\bigl( (W_0)^4 \hat{q}^{L_0-\frac{c}{24}}\bigr) 
\approx \frac{\alpha^4\tau^8}{4!} \int F\big( (W,z_1), (W,z_2), (W,z_3), (W,z_4); \tau\big) \ ,
\ee
where throughout this subsection,
\be
\int = \int_{1}^q \frac{dz_1}{z_1} \cdots \int_1^q\frac{dz_4}{z_4} \ .
\ee
Applying the recursion relation (\ref{recur}), we get
\begin{align}\label{four1}
\int F\big((W,z_1),\dots,(W,z_4); \tau\big) = & \int F\big(W_0; (W, z_2), (W,z_3), (W, z_4); \tau\big) \\
& + 3 \int {\cal P}_{\ell+1}\left(z_4\over z_1\right) F\big((W, z_2), (W,z_3),(W[\ell]W,z_4);\tau \big)\ ,\notag
\end{align}
where sums over Weierstrass indices are always implied.  Three terms on the right have been combined by relabelling integration variables; strictly speaking, this is not allowed, because the integration contours are displaced by phases $\varepsilon_j$ as in figure \ref{fig:contours}, but we will account for this subtlety below. Note that we cannot drop the first term on the right, despite the zero mode which naively implies that the leading $\mathcal{O}(q^{-c/24})$ contribution vanishes. The reason is that $q$ also appears in the limits of integration, and after integration we will see that this term contributes at leading order.

\noindent
Applying the recursion relation twice more to (\ref{four1}) gives
\begin{align}
\int F\big((W,z_1),&\dots,(W,z_4);\tau\big)=  \nonumber \\ 
& 3 \int \wP_{\ell+1}\left(z_4\over z_1\right) \wP_{m+1}\left(z_4\over z_2\right) 
\wP_{k+1}\left(z_4\over z_3\right)\langle W[k]W[m]W[\ell]W_{-3}\rangle  
 \nonumber   \\
& +3  \int {\cal P}_{\ell+1}\left(z_4\over z_1\right) \wP_{m+1}\left(z_3\over z_2\right) \wP_{k+1}\left(z_4\over z_3\right) \langle (W[m]W)[k]W[\ell]W_{-3}\rangle
\nonumber \\
& + \frac{5 (2 \pi i)}{m} \int \wP_{\ell+1}\left(z_4\over z_1\right) \p_\tau \wP_{m}\left(z_4\over z_3\right) \langle d^{(1)}[m]W[\ell]W_{-3}\rangle 
\label{thirdapp} \\
& + \frac{2 (2 \pi i)}{\ell} \int \p_\tau \wP_{\ell}\left(z_4\over z_2\right) \wP_{m+1}\left(z_4\over z_3\right)\langle W[m]d^{(1)}[\ell]W_{-3}\rangle 
\nonumber
 \\
& + \frac{(2 \pi i)^2}{\ell(\ell-1)}\int \p_\tau^2 
\wP_{\ell-1}\left(z_4\over z_3\right)\langle d^{(2)}[\ell]W_{-3}\rangle\ ,  \nonumber
\end{align}
where $\langle \cdot \rangle$ is the connected part of the vacuum expectation value and we have defined the states
\be
d^{(1)} = - W[0]W \ , \qquad \hbox{and} \qquad d^{(2)} = W[0]W[0]W \ .
\ee
Each of these integrals can be computed easily using properties of the Weierstrass functions, see appendix \ref{sec:Weierstrass} for details and examples.  Some of the integrals diverge if we ignore the phases $\varepsilon_j$ that separate the integration contours.  For example, making the contour separations explicit, the last line includes
\be
\tau^2 \int_{\varepsilon_3}^{q\varepsilon_3}\frac{dz_3}{z_3}
\int_{\varepsilon_4}^{q\varepsilon_4}\frac{dz_4}{z_4}\, \,
\p_\tau^2 {\cal P}_2\left(z_4\over z_3\right) \propto 
\tau^2  {\cal P}_2\left(\varepsilon_4\over\varepsilon_3\right) \ ,
\ee
which diverges as $\varepsilon_4 \rightarrow \varepsilon_3$. However, since the original expression is independent of the contours, we can account for all of the contour dependence by  simply dropping divergent terms of the form ${\cal P}_{m}(1)$.  These terms must cancel in the sum, and the pole structure (as functions of the $\varepsilon_j$) requires that they cancel only with other Weierstrass functions of the same order, so they do not leave any finite remainder.\footnote{As a check, we have confirmed by explicit computation that the contour-dependent terms in the 4-point recursion cancel in the final answer.  This requires keeping track of contours at each step in the recursion and computing many additional terms.}

The same argument implies that we can drop terms proportional to $\tau^2$ or higher powers of $\tau$.  As noted in section \ref{sec:strategy}, the powers of $\tau$ and $\alpha$ in the partition function (\ref{gravres}) are dictated by conformal invariance.  Only the $\tau^1$ term in (\ref{thirdapp}) has the correct power; other powers can only appear when multiplied by appropriate $\varepsilon_j$-dependent functions (or suppressed by powers of $q$), and such terms cannot contribute.

As a consequence, only terms with bracketed indices summing to 3 need to be computed in (\ref{thirdapp}). That is, 
\be\label{z4ans}
\int F\big((W,z_1),\dots,(W,z_4);\tau\big) = 3I_1(A_1 + A_2) + 3 I_2 A_6 
+ \frac{5}{2}I_3A_3 + \frac{5}{3}I_4 A_4 + I_3 A_5 \ ,
\ee
where the integrals are
\begin{align}\label{intlist}
I_1 &= \frac{1}{(2\pi i)^{6}}\!\int \P{2}{z_4}{z_1}\P{2}{z_4}{z_2}\P{2}{z_4}{z_3} &\notag
I_2 &= \frac{1}{(2\pi i)^{6}}\!\int \P{1}{z_4}{z_1}\P{1}{z_3}{z_2}\P{4}{z_4}{z_3} \\
I_3 &= \frac{1}{(2\pi i)^{5}}\!\int \P{2}{z_4}{z_1}\p_\tau \P{2}{z_4}{z_3} & 
I_4 &= \frac{1}{(2\pi i)^{5}}\!\int \P{1}{z_4}{z_1}\p_\tau \P{3}{z_4}{z_3} \ ,
\end{align}
and the algebra factors are
\begin{align}\label{afactors}
A_1 &= (2\pi i)^6 \langle W[1]W[1]W[1]W_{-3} \rangle & 
A_2 &= (2\pi i)^6\langle (W[1]W)[1] W[1] W_{-3} \rangle \notag \\
A_3 &= (2\pi i)^6\langle d^{(1)}[2]W[1]W_{-3}\rangle & 
A_4 &= (2\pi i)^6\langle d^{(1)}[3]W[0]W_{-3}\rangle \\
A_5 &= (2\pi i)^6\langle W[1]d^{(1)}[2]W_{-3}\rangle &
A_6 &= (2\pi i)^6  \langle (W[0]W)[3]W[0]W_{-3}\rangle  = - A_4 \ . \notag
\end{align}
Here the identity between $A_4$ and $A_6$ follows from the definition of $d^{(1)}$. 
To compute the vevs in (\ref{afactors}), it suffices to ignore nonlinear terms in the $\W$-algebra, as they contribute at subleading order in $1/c$ (though this will no longer be the case in the six-point function). Furthermore, we discard disconnected contributions, as described above. Nested bracket modes like $(W[1]W)[1]$ can be computed directly by evaluating the state $W[1]W_{-3}\Omega$, computing the corresponding $[1]$-mode, etc.  A perhaps simpler approach, especially for higher $n$-point functions where nonlinear composite states contribute, is to use the Jacobi identity (see \cite{Zhu})
\be\label{jacobi1}
(a[m]b)[n] = \sum_{i=0}^\infty
\left[(-1)^i{m \choose i}a[m-i]b[n+i] - (-1)^{m+i}{m \choose i}b[m+n-i]a[i]\right] \ ,
\ee
to remove the nested modes. Applying this identity, plugging in the definition of the bracket modes (\ref{abradef}), and using the $\w{\lambda}$ commutators, we find
\begin{align}
A_1 &= c\left(-\frac{N_3^2}{9} + \frac{26 N_4}{135}\right) &
A_2 &=  c\left(-\frac{4N_3^2}{27} + \frac{2N_4}{15}\right) 
\end{align}
and
\be
A_3 = A_2 \ , \qquad
A_4 = -A_6 = \frac{3}{2}A_2 \ , \qquad
A_5 = A_1 \ . 
\ee
The Weierstrass integrals $I_{1,2,3,4}$ are computed in appendix \ref{sec:Weierstrass}. Plugging 
in the answers it then finally follows from (\ref{z4}) and (\ref{z4ans}) that 
\bea
Z^{(4)} &\approx&  -q^{-c/24}\, 2\pi i c \, \frac{2}{27}
\left(5N_3^2 - 7 N_4\right)\alpha^4\tau^9\\
&\approx& -q^{-c/24}\, 2\pi i c \, \frac{50}{81}\, 
\frac{\lambda^2-7}{\lambda^2-4}\alpha^4\tau^9
\eea
in precise agreement with the gravity answer (\ref{gravcon}).

\subsection{The 6-point function}\label{ss:sixpoint}

A new ingredient appears in the $\alpha^6$ correction to the partition function: nonlinear terms in the algebra contribute at leading order.  Schematically, the spin-3 and spin-4 commutators of $\w{\lambda}$ have the form
\begin{equation}
\begin{array}{rcl}
\ [W, W] &\sim& {\displaystyle U + L + \frac{1}{c}L^2 + c } \\[5pt]
\ [U, W] &\sim& {\displaystyle X + W + \frac{1}{c}LW}  \ ,
\end{array}
\end{equation}
where $U$ is the spin-4 current and $X$ is the spin-5 current (see appendix \ref{a:commutators} 
for details). Note that nonlinear terms always appear with powers of $1/c$, which is what 
allowed us to drop them in the 2- and 4-point calculation, but starting from the 6-point case 
they can nonetheless contribute.  For example, starting with the the 6-point function we can contract two currents to make $L$ and two currents to make $U$,
\be
\bcontraction[1ex]{}{W}{}{W} WW
\bcontraction[1ex]{}{W}{}{W} WW
WW \rightarrow LUWW \ .
\ee
Now contracting $\bcontraction[1ex]{}{U}{}{W} UW \sim \frac{1}{c}LW + X + W$ gives the nonlinear term
\be
\frac{1}{c}LLWW \ .
\ee
Since $LL$ and $WW$ both have central terms, this term is of order $c$, and therefore contributes to $\log Z$ at leading order. 

Otherwise, the computation of the 6-point function proceeds as above. Starting from
\be\label{z6}
Z^{(6)} \equiv \frac{(2\pi i\alpha)^6}{6!} \,
\Tr \bigl( (W_0)^6\hat{q}^{L_0-\frac{c}{24}} \bigr) \approx \frac{\alpha^6\tau^{12}}{6!}
\int F\big((W,z_1),\dots,(W,z_6); \tau\big) \ ,
\ee
we apply the recursion relation to reduce the six-point correlator to a sum of vevs, perform the Weierstrass integrals, discard contour-dependent terms (i.e.\ ${\cal P}_{m}(1)$ and 
terms not linear in $\tau$), and compute the sum.  Some additional details are provided in appendix \ref{a:sixpoint}.  The final result is
\bea
Z^{(6)} &\approx& q^{-c/24}\, 2\pi i c\,  \left(\frac{17 N_3^3}{648} - \frac{581N_3N_4}{9720} + \frac{497N_4^2}{12150N_3} + \frac{101N_5}{2160}\right) \alpha^6 \tau^{13}\\
&\approx& q^{-c/24}\, 2\pi i c\, \frac{200}{81}\, \frac{5\lambda^4-85\lambda^2 + 377}{(\lambda^2-4)^2}\alpha^6\tau^{13}
\eea
in precise agreement with the black hole value (\ref{gravcon}).

\section{Discussion}\label{s:discussion}

As we have shown in this paper, we can reproduce the higher spin corrections to the
black hole entropy from integrals of correlation functions of $\W$-currents on the torus.
This was to be expected on general grounds, but it is satisfying to see that things work
out correctly. In particular, the CFT calculation depends on the full non-linear 
$\W_\infty[\lambda]$ structure and hence our result provides a detailed check that 
$\W_\infty[\lambda]$
is indeed the correct symmetry algebra of the dual CFT. We should also stress that, on
the face of it, our computation is entirely different from the bulk calculation 
performed in \cite{Kraus:2011ds}.  In the bulk, the entropy is determined by the solution of a 
zero-holonomy condition for \hs{\lambda} Chern-Simons gauge fields on a torus.  The 
zero-holonomy condition ensures smoothness of the Euclidean horizon, which determines 
the thermodynamics of the black hole.\footnote{Using this method, the authors of 
\cite{Kraus:2011ds}  were able to compute also the $\alpha^8$ term in (\ref{gravres}), whereas 
to do so in CFT by our method would be very difficult.}  Somehow, the holonomy condition encodes the full nonlinear structure of $\w{\lambda}$ in a nice geometric way; it would be very interesting to understand how the Chern-Simons calculation emerges directly from CFT or vice versa.

Let us close by commenting on the regime of  validity for the free energy formula 
(\ref{gravres}) and its interpretation in higher spin holography.  In any unitary CFT 
with \w{\lambda} symmetry, assuming the theory has a gap, the formula applies 
asymptotically as $T \rightarrow \infty$.  However, this fact alone is not enough to 
conclude that a CFT with \w{\lambda} symmetry reproduces the thermodynamics of 
higher spin black holes.  Indeed {\em any} CFT obeys the Cardy formula asymptotically, 
but not every CFT describes black holes in AdS$_3$.  
One key feature that distinguishes holographic CFTs from the rest is that the Cardy formula 
applies not only asymptotically, but at temperatures down to the Hawking-Page transition in the 
bulk.  These are very special CFTs, like the D1-D5 system dual to string theory in AdS$_3$, 
with a large gap in dimensions separating excited string states from the low-energy fields of 
Einstein gravity, see e.g.\  \cite{Dijkgraaf:2000fq}.

For the case at hand, the dual CFT 
must have free energy (\ref{gravres}) whenever the black hole dominates the bulk 
thermodynamics.  However, it is not known whether Vasiliev gravity in three 
dimensions has a Hawking-Page transition, or whether the black hole dominates the bulk thermodynamics anywhere besides $T \rightarrow \infty$.  
If there is indeed a phase transition above which the black hole dominates, then the dual 
CFT should have a gap large enough so that (\ref{gravres}) applies above the transition 
temperature. The 
microscopic CFT proposed in \cite{Gaberdiel:2010pz}, a $\W_N$ minimal model, has a large 
number of light states with dimension $\Delta < 1$, so it presumably obeys (\ref{gravres}) 
only at asymptotically high temperatures. This may be mirrored by the fact that the Vasiliev gravity theory 
has other,  black-hole-like solutions \cite{Castro:2011iw} which contribute to the bulk thermodynamics.

\bigskip
\textbf{Acknowledgments}

We thank Antal Jevicki, Per Kraus, Jan Manschot, Steve Shenker and Cristian Vergu for discussions. TH is grateful to the Stanford Institute for Theoretical Physics for its hospitality during a portion of this work, and acknowledges support by the U.S. Department of Energy, grant DE-FG02-90ER40542. The research of MRG and KJ is supported in part by the Swiss National Science Foundation. MRG thanks the Aspen Center of physics for hospitality during the initial stages of this project. TH and MRG
thank the Isaac Newton Institute in Cambridge for hospitality during the final stages of this work.

\settocdepth{section}
\appendix

\section{Weierstrass functions}\label{sec:Weierstrass}

The Weierstrass functions are defined by the power series 
\be\label{defp}
{\cal P}_k(x,q) = \frac{(2\pi i)^k}{(k-1)!} \sum_{n\neq 0} \left(
\frac{n^{k-1}x^n}{1-q^n} \right) \ ,
\ee
for $k \geq 1$, which converge for $|q|<|x|<1$. (This differs slightly from the standard Weierstrass function $\wp_k$, see \cite{Zhu} for the relation.) They satisfy the important recursion relation
\be\label{Prec}
x \frac{d}{dx} {\cal P}_k(x,q) = \frac{k}{2\pi i} {\cal P}_{k+1}(x,q) \ .
\ee
We can rewrite ${\cal P}_1(x,q)$ as 
\be
{\cal P}_1(x,q) = \frac{2\pi i}{1-x} - 2\pi i +  2\pi i
\sum_{n\neq 0} \left( \frac{x^n q^n }{1-q^n}\right) \ ,
\ee
thus defining a meromorphic continuation of ${\cal P}_1$ 
(and hence, because of (\ref{Prec}), for all ${\cal P}_k$) to $|q|<|x|<|q|^{-1}$.
A straightforward calculation then shows the
identity
\be
{\cal P}_1(qx,q)={\cal P}_1(x,q)+2\pi i\ , \qquad
{\cal P}_k(qx,q)={\cal P}_k(x,q) \quad (k>1) \ . \label{Pperiod}
\ee
In the recursion formula (\ref{multrec}), we also need the closely related functions
\be\label{defg}
g^i_k(x,q) \equiv \frac{(2\pi i)^{k}}{(k-1)!} \sum_{n\neq 0} n^{k-i-1}x^n \p_\tau^i\left(1\over 1-q^n\right) \ .
\ee
Note the relations
\be
g^0_k(x,q) =\, {\cal P}_k(x,q)
\ee
and 
\be\label{gderivap}
g^i_k(x,q) = (2\pi i)^i \frac{(k-i-1)!}{(k-1)!}\p_\tau^i {\cal P}_{k-i}(x,q) \quad (k-i \geq 1) \ .
\ee

\subsection{Weierstrass integrals}

Products of ${\cal P}_k$ and $g^i_k$ can be integrated using (\ref{Prec}) and (\ref{gderivap}). 
Dropping from now on the argument $q$, we have for example from (\ref{Prec}) and 
(\ref{Pperiod}),
\be
\int_1^q \frac{dz_2}{z_2}\, 
{\cal P}_2\left(z_1\over z_2\right) = (2\pi i)^2 \  , \quad 
\int_1^q \frac{dz_2}{z_2}\, {\cal P}_{m+1}\left(z_1\over z_2\right) = 0 \quad (m>1) \ .
\ee
The integral of ${\cal P}_1$ can be computed by relating ${\cal P}_1$ to the Weierstrass $\zeta$-function (denoted $\wp_1$ in \cite{Zhu}), which integrates to the log of the Weierstrass $\sigma$-function, giving  
\be
\int_1^q\frac{dz_2}{z_2}\, \P{1}{z_1}{z_2} = (2\pi i) \left(i\pi-2\pi i \tau + \log z_1\right)  \ .
\ee
Other useful identities can be derived by integrating by parts, for example
\be
\int_{1}^{q}\frac{dz_1}{z_1} \, \log z_1 \, \P{m+1}{z_1}{z_2} 
= (2\pi i)^2\,\frac{ \tau}{m}\, \P{m}{1}{z_2} - (2\pi i)^3\, \frac{1}{m} \delta_{m,2} \quad (m\geq 2) \ .
\ee
We will also need to integrate $g^i_k(x,q)$.  In all of our computations, $k \geq i+1$, so we can replace $g^i_k$ (up to a constant factor) 
by $\p_\tau^i {\cal P}_{k+1-i}$.  Integrals of $\p_\tau^i {\cal P}_m$ can be computed by 
differentiating the identities above.  For example,
\be\label{dtauid}
\p_\tau \int_1^{q}\frac{dz_2}{z_2}\, \P{m}{z_1}{z_2} = 0 \quad(m>1)
\ee
implies, acting with the derivative on the limits of integration and on the integrand,
\be\label{tauint}
\int_1^q\frac{dz_2}{z_2}\, \p_\tau \P{m}{z_1}{z_2} = - (2\pi i)\, \mathcal{P}_m (z_1) \quad (m>1) \ .
\ee
Differentiating again, 
\be
\int_1^q\frac{dz_2}{z_2}\, \p_\tau^2\P{m}{z_1}{z_2} = -(2\pi i)\, 2\, \p_\tau \mathcal{P}_m (z_1) 
- (2\pi i)\, m \mathcal{P}_{m+1} (z_1)  \quad (m>1) \ .
\ee
Combining these techniques allows one to compute all of the integrals encountered in the partition function calculations.  For the four-point function, the relevant integrals from (\ref{thirdapp}) are listed in (\ref{intlist}).
$I_1$ is computed by integrating first over $z_1,z_2,z_3$, then over $z_4$, which immediately gives
\be
I_1 = 2\pi i \tau \ .
\ee
To compute $I_2$, we first integrate over $z_1,z_2$, and then integrate by parts to get 
\be
I_2 = \frac{1}{6}\, (2\pi i \tau) - \frac{1}{6} \tau^2 \, \mathcal{P}_2 (1) \quad 
\rightarrow \quad \frac{1}{6}\, (2 \pi i \tau) \ ,
\ee
where in the last step we have used that the divergent term ${\cal P}_2(1)$ cancels if one carefully accounts for contour separations (see section \ref{ss:fourpoint}).

For $I_3$, we integrate first over $z_1$ then apply (\ref{tauint}), resulting in
\be
I_3 = - 2\pi i \tau \ .
\ee
Similarly $I_4$ is determined by first integrating over $z_1,z_3$, and then integration by parts, leading to 
\be
I_4 = \frac{1}{2}\, (2\pi i \tau) - \frac{1}{2} \tau^2 \, \mathcal{P}_2(1) \quad
\rightarrow  \quad \frac{1}{2}\, (2 \pi i \tau) \ .
\ee
All of the integrals encountered in the 6-point function can be computed similarly.

\section{Torus recursion}\label{sec:recursion}

\subsection{Bracketed modes}\label{sec:brecketed}

Let us define the expansion
coefficients $c(h,j,m)$ via
\be
\bigl( \log(1+z) \bigr)^s\, (1+z)^{h-1} = \sum_{j\geq s} c(h,j,s)\, z^j \ .
\ee
Then the bracketed modes can be defined using the `Mathematician' convention for modes
as
\be\label{abradef0}
a[s] = (2\pi i )^{-s-1}\, \sum_{i=s}^{\infty} c(h_a,i,s) \, a(i) \ . 
\ee
Here we have assumed that $a$ has definite $L_0$ eigenvalue, $L_0 a = h_a a$, and 
the modes $a(i)$ are related to the modes $a_i$ via 
\be\label{rel1}
a(n) = a_{n+1-h_a}
\ee
since we have 
\be\label{rel2}
V(a,z) = \sum_{n} a(n) z^{-n-1} = \sum_n a_n z^{-n-h_a} = \sum_n a_{n+1-h_a} z^{-n-1} \ .
\ee
Thus in terms of the usual physicists' modes we can rewrite (\ref{abradef0}) as
\be\label{abradef}
a[s] = (2\pi i )^{-s-1}\, \sum_{j=s+1-h_a}^{\infty} c(h_a,j+h_a-1,s)\, a_j \ . 
\ee
For the Virasoro modes a different convention is usually used, namely
\be\label{Ldef}
L_{[n]} = (2\pi i )^{-n}  \sum_{j\geq n+1} c(2,j,n+1) L_{j-1} - (2\pi i)^2 \frac{c}{24} \delta_{n,-2} \ .
\ee
The relation to the above definition is 
$L_{[n]} = (2\pi i)^2 (\omega-\tfrac{c}{24} {\bf 1})[n+1]$, where $\omega$ is the field
with mode expansion
\be
V(\omega,z) = \sum_{n} \omega(n) z^{-n-1} = \sum_n L_n z^{-n-2}\ .
\ee
Note that $(\omega-\tfrac{c}{24} {\bf 1})$ has definite $L_{[0]}$ eigenvalue since 
\be
L_{[0]} = L_0 + \tfrac{1}{2} L_1 - \tfrac{1}{6} L_2 + \cdots 
\ee
and hence
\be
L_{[0]} (L_{-2} \Omega -  \tfrac{c}{24} \Omega) = 2 L_{-2} \Omega - \tfrac{1}{6} L_2 L_{-2} \Omega
= 2 \bigl( L_{-2} \Omega - \tfrac{c}{24} \Omega\bigr) \ .
\ee
\smallskip

\noindent In these conventions we then have the identity 
\be\label{iden}
\bigl( L_{[-1]} a \bigr) [n] = - n a[n-1] \ , 
\ee
which becomes for states of definite $L_{[0]}$ eigenvalue the identity 
$(L_{[-1]}a)_{[n]} = - ([h_a]+n) a_{[n]}$, see page 4 of \cite{Gaberdiel:2008pr}. 

\subsection{Derivation of recursion relations}

In this appendix we prove the recursion relation (\ref{multrec}).  The case without zero modes is derived in \cite{Zhu} and reviewed in \cite{Gaberdiel:2008pr}, and a similar strategy is followed here.  We begin by expanding the first vertex operator in modes,
\begin{align}
F\big(b_0^\ell; (a^1,z_1),\dots,(a^n,z_n); \tau\big) &= F
\left(b_0^\ell\, a_0; (a^2,z_2),\dots,(a^n,z_n); \tau \right) \\
& \quad + \sum_{k\neq 0}z_1^{-k} F\left(b_0^\ell a^1_k; (a^2,z_2),\dots,(a^n,z_n); \tau\right)\ . \notag
\end{align}
We will now cycle $a^1_k$ around the trace to evaluate the second line.  For this we need the commutator
\be\label{vertexcom}
[a^1_k, V(a^j,z_j)] = \sum_{m=0}^{\infty}\, {h_1 - 1 + k\choose m}\, 
V(a^1_{m-h_1+1}a^j,z_j)\, z_j^{h_1-1+k-m} \ ,
\ee
and the relation
\be\label{pastq}
a^1_k\, q^{L_0-\frac{c}{24}} = q^k \, q^{L_0-\frac{c}{24}}\, a_k^1 \ .
\ee
When we move $a^1_k$ past the zero modes $b_0^\ell$ we will pick up the commutator 
$[a^1_k, b_0^\ell]$.  Let us define the modes of the single commutator to be 
$d_k^{(1)} = - \tfrac{1}{2\pi i}[b_0, a^1_k]$, and more generally,
\begin{equation}
d_k^{(0)} = a^1_k \ , \qquad\qquad  d_k^{(i)} = -\frac{1}{2\pi i} [b_0, d_k^{(i-1)}] \qquad (i>0) \ .
\end{equation}
This definition can be recast in terms of states 
\be
d^{(i)} = (-1)^i\,  (b[0])^i a^1 \ ,
\ee
as follows from (\ref{zeroco}). Next we prove  the identity
\begin{align}\label{recursionid}
z_1^{-k}F\big(&b_0^\ell a^1_k; (a^2,z_2), \dots, (a^n, z_n); \tau\big) = \\
&  \sum_{j=2}^n\sum_{i=0}^\ell {\ell \choose i}
k^{-i}\left(\frac{z_j}{z_1}\right)^{k}\p_\tau^i\left(\frac{1}{1-q^k}\right) \notag\\
& \quad \times \sum_{t=0}^{\infty} {h-1+k\choose t} 
F\left(b_0^{\ell-i}; (a^2,z_2), \dots, (d^{(i)}_{t-h+1}a^j,z_j),\dots,(a^n,z_n); \tau\right) \ ,\notag
\end{align}
where in the last line, $h$ is the dimension of $d^{(i)}$. (Typically $d^{(i)}$ is not homogeneous under $L_0$, in which case this expression is defined by expanding in $L_0$ eigenstates.)  This is proved by induction on 
$\ell$.  Moving $a^1_k$ around the trace on the l.h.s., we pick up a commutator with each vertex operator and a commutator with $b_0^\ell$.  When it returns to its original position, the trace is multiplied by $q^k$ from (\ref{pastq}).  Solving for the original expression,
\begin{align}
z_1^{-k}F\big(& b_0^\ell a^1_k; (a^2,z_2), \dots, (a^n,z_n); \tau\big) = \\
& \frac{1}{1-q^k}\sum_{j=2}^n \left(\frac{z_j}{z_1}\right)^{k}
\sum_{t=0}^{\infty}{h_1-1+k\choose t}F\left(b_0^\ell; (a^2,z_2), \dots, (a^1_{t-h_1+1}a^j,z_j),\dots,(a^n,z_n); \tau\right)\notag\\
& - z_1^{-k}\frac{q^k}{1-q^k} F\left( [b_0^\ell, a_k^1]; (a^2,z_2), \dots, (a^n,z_n);\tau\right)\ .\notag
\end{align}
The second line is the $i=0$ term in (\ref{recursionid}), which proves the identity for $\ell=0$. To evaluate the third line, we plug in the commutator identity
\be
[b_0^{\ell}, a_k^1] = -\sum_{s=1}^{\ell}(2\pi i)^s {\ell \choose s}\, b_0^{\ell-s}d_k^{(s)} \ ,
\ee
and apply the induction hypothesis.  Finally, we trade $i$ for $p=i+s$ and use the identity
\be
\sum_{s=1}^p {\ell \choose s}{\ell-s \choose p-s}\frac{q^k}{1-q^k}k^{-p+s}(q\p_q)^{p-s}\frac{1}{1-q^k} = {\ell \choose p}k^{-p}(q\p_q)^p\frac{1}{1-q^k} 
\ee
to simplify the result, proving (\ref{recursionid}).

The next step is to rewrite (\ref{recursionid}) using bracket modes and Weierstrass functions.  The key relation is
\be
\sum_{t= 0}^{\infty}\sum_{k\neq 0}{h_a - 1 + k\choose t}\frac{1}{1-q^k}\, x^k \, a_{t-h_a+1} 
=  \sum_{m=0}^{\infty}{\cal P}_{m+1}(x,q) \, a[m] \ ,
\ee
which is straightforward to derive by expanding and rearranging the sums.  Differentiating with respect to $\tau$ and integrating with respect to $x$, this also implies
\be
\sum_{t=0}^{\infty}\sum_{k\neq 0}{h_a - 1 + k\choose t}k^{-i}\p_\tau^i\left(\frac{1}{1-q^k}\right)x^k a_{t-h_a+1} = \sum_{m=0}^{\infty}g^i_{m+1}(x,q)\, a[m] \ ,
\ee
where $g^i_{m+1}(x,q)$ is defined in appendix \ref{sec:Weierstrass}. Setting $x=z_j/z_1$ and summing over $k$ in (\ref{recursionid}) completes the proof of the recursion relation (\ref{multrec}).

\section{Details of the 6-point calculation}\label{a:sixpoint}

In this appendix we provide additional technical details on the $\alpha^6$ correction to the partition function outlined in section \ref{ss:sixpoint}. Define
\bea
\II{\ell m k r}{\mu \nu \rho\gamma}{\sigma\alpha\beta\delta}&=& \frac{2 \pi i}{r}\int_{1}^q\frac{dz_1}{z_1}\cdots\! \int_{1}^q\frac{dz_6}{z_6}\, 
\P{\ell+1}{z_\mu}{z_\sigma}\P{m+1}{z_\nu}{z_\alpha}\P{k+1}{z_\rho}{z_\beta}\p_\tau \P{r}{z_\gamma}{z_\delta}\\
\II{\ell m k r j}{\mu\nu\rho\gamma\chi}{\sigma\alpha\beta\delta\psi} &=& \int_{1}^q\frac{dz_1}{z_1}\cdots\! \int_{1}^q\frac{dz_6}{z_6}\,
\P{\ell+1}{z_\mu}{z_\sigma}\P{m+1}{z_\nu}{z_\alpha}\P{k+1}{z_\rho}{z_\beta} \P{r+1}{z_\gamma}{z_\delta}\P{j+1}{z_\chi}{z_\psi} \ .\notag
\eea
Applying the recursion relation (\ref{multrec}) repeatedly to (\ref{z6}) gives
\begin{align}\label{sixpointbig}
\int F\big((W,z_1),&\dots,(W,z_6);\tau\big) =\\
& 75 \langle (W[\ell]W)[r]d[k]W[m]W_{-3}\rangle \II{\ell m r k}{6555}{2364}\notag\\
& + 19\langle d[r]W[k]W[m]W[\ell]W_{-3}\rangle \II{\ell m k r}{6656}{2345}\notag\\
& + 120\langle (W[\ell]W)[r]W[0]W[k]W[m]W_{-3}\rangle
\II{\ell m k r}{6555}{2346}\notag\\
 & + 14 \langle W[r]d[k]W[m]W[\ell]W_{-3}\rangle \II{\ell m r k}{6666}{2354}\notag\\
 & + 40 \langle(W[m]W[\ell]W)[r]d[k]W_{-3}\rangle \II{\ell m r k }{6655}{2364}\notag\\
 & + 30 \langle (W[\ell]W)[r]W[k]d[m]W_{-3}\rangle \II{\ell k r m}{6555}{2463}\notag\\
 & + 9 \langle W[r]W[k]d[m]W[\ell]W_{-3}\rangle \II{\ell k r m}{6666}{2453}\notag\\
 & + 4 \langle W[r]W[k]W[m]d[\ell]W_{-3}\rangle \II{mkr\ell}{6666}{3452}\notag\\
 & + 45 \langle (W[\ell]W)[j]W[r]W[k]W[m]W_{-3}\rangle \II{\ell m k r j}{63333}{12546}\notag\\
 & + 40 \langle (W[k]W[m]W)[j]W[r]W[\ell]W_{-3}\rangle \II{\ell m k r j}{63363}{12546} \notag\\
 & + 5 \langle W[j]W[r]W[k]W[m]W[\ell]W_{-3}\rangle \II{\ell m k r j}{66666}{12345}\notag\\
 & + 15 \langle (W[k]W)[r](W[m]W)[0]W[\ell]W_{-3}\rangle \II{\ell m k r}{6346}{1254}\notag\\
 & + 30 \langle (W[\ell]W)[j](W[m]W)[r]W[k]W_{-3}\rangle\II{\ell m k r j}{63444}{12536} \ ,\notag
\end{align}
where $d = -W[0]W$. We have discarded terms not linear in $\tau$ because all such terms must cancel by the argument in section \ref{ss:fourpoint}. The vevs are computed by applying the Jacobi identity (\ref{jacobi1}) to remove nested bracket modes.  Only the terms with bracket mode indices adding up to 5 will integrate to be finite, contour-independent contributions, so all others cancel. There are 37 non-zero contractions of the form
\be
\langle W[i]W[j]W[k]W[l]W[m]W_{-3}\rangle \ ,
\ee
satisfying $i+j+k+l+m=5$. These are computed with the help of  Mathematica by applying the commutation relations and expanding out composite currents in products of modes, giving for example,
\be
\langle W[1]^5 W_{-3}\rangle = (2\pi i)^{-10}c\left(\frac{10N_3^3}{9} - \frac{364N_3N_4}{135} + \frac{2704N_4^2}{2025N_3} + \frac{179N_5}{63}\right) \ ,
\ee
and other similar formulae. Each of the corresponding integrals can be computed straightforwardly using the methods described in appendix \ref{sec:Weierstrass}.

\section{The \w{\lambda} algebra}\label{a:commutators}

Following \cite{Gaberdiel:2011wb} (see also \cite{Campoleoni:2011hg}), but including some higher orders, the first few commutation relations of the the nonlinear $\w{\lambda}$ algebra are
\begin{align}\label{modecom}
[L_m, L_n] &= (m-n)L_{m+n} + \frac{c}{12}m(m^2-1)\delta_{m,-n}\\
[L_m, W_n] &= (2m-n)W_{m+n}\notag\\
[W_m, W_n] &= 2(m-n)U_{m+n}+  \frac{N_3}{12}(m-n)(2m^2+2n^2-mn-8)L_{m+n}\notag \\ & \ \ \ \ + \frac{N_3 c}{144}m(m^2-1)(m^2-4)\delta_{m,-n} + \frac{8 N_3}{c}(m-n)\Lambda^{(4)}_{m+n} \notag\\
[W_m, U_n] &= (3m-2n)X_{m+n} + \frac{N_4}{15 N_3}(n^3 - 5m^3 - 3 m n^2 + 5 m^2 n - 9n + 17 m)W_{m+n}\notag \\
& \ \ \ \ - \frac{208 N_4}{25 c N_3}(3m-2n)\Lambda^{(5)}_{m+n} - \frac{84N_4}{25 c N_3}\Theta^{(6)}_{m+n} \notag\\
[U_m, U_n] &= 3(m-n)Y_{m+n} -(m-n) \frac{ N_4 n_q }{cN_3^2} \Lambda^{(6)}_{m+n}
+ n_{44} (m-n)(m^2-mn+n^2-7)U_{m+n} 
 \notag\\
&\ \ \ \ -\frac{N_4}{360}(108 - 39 m^2 + 3 m^4 + 20 m n - 2 m^3 n - 39 n^2+ 4 m^2 n^2 - 
 2 m n^3 + 3 n^4) \notag\\
& \ \ \ \ \ \ \ \  \times (m-n)L_{m+n}  
- \frac{c N_4}{4320} m (m^2-1)(m^2-4)(m^2-9) \delta_{m,-n} +\cdots \notag\\
[W_m, X_n] &= (4m-2n)Y_{m+n} - \frac{1}{56}\frac{N_5}{N_4}(28m^3-21m^2n + 9 mn^2 - 2n^3 - 88 m + 32 n)U_{m+n}\notag\\
& \ \ \ \ +\frac{42  N_5}{5 cN_3^2}(2m-n)\Lambda^{(6)}_{m+n} + \cdots\notag \\
[U_m, X_n] &= (4m-3n)Z_{m+n} + n_{45} (14 m^3-5 n^3-21 m^2 n+15 m n^2-86 m+59 n) X_{m+n} \cr
&\ \ \ \ +\frac{N_5}{1680 N_3}(28 m^5-3 n^5-35 m^4 n+30 m^3 n^2-20 m^2 n^3+10 m n^4 \cr
&\ \ \ \ \ \ \ \ -340 m^3+355 m^2 n-220 m n^2+75 n^3+792 m-432 n) W_{m+n}+\cdots\notag\\
[X_m, X_n] &= \frac{ cN_5}{241920}m(m^2-1)(m^2-4)(m^2-9)(m^2-16)\delta_{m,-n} + \cdots \ ,\notag
\end{align}
where $L,W,U,X,Y,Z$ are the currents of spin $2,3,4,5,6,7$ respectively, and the constants are
\begin{align}
N_3 &=  \frac{16}{5}\, \sigma^2 \, (\lambda^2-4)  & N_4 &=  - \frac{384}{35}\, \sigma^4 \, (\lambda^2-4)\,  (\lambda^2-9) \notag\\
N_5 &=  \frac{4096}{105}\, \sigma^6\, (\lambda^2-4)  (\lambda^2-9) \,  (\lambda^2-16) \ , & n_{44} &= \frac{8}{15}\sigma^2(\lambda^2-19)\notag\\
n_{45} &= \frac{8}{385}\sigma^2(3\lambda^2-97) & n_q &= \frac{32}{10}\sigma^2(29\lambda^2-284)\notag \ .
\end{align} 
Here $\sigma$ is an arbitrary choice of normalization. We use the normalization of \cite{Kraus:2011ds}, in which the $WW$ OPE is
\be
W(z) W(0) \sim  \frac{10c}{3}\frac{1}{z^6} + \cdots \ .
\ee
This corresponds to
\be
\sigma^2 = \frac{5}{4(\lambda^2-4)} \ .
\ee
(For comparison to \cite{Kraus:2011ds}, note that the charges are related by $W_0^{\rm here} = 2\pi {\cal W}_0^{\rm there}$; for currents, however, $W^{\rm here} = 2\pi i {\cal W}^{\rm there}$, because the zero-mode in \cite{Kraus:2011ds} is defined on the cylinder whereas ours is defined on the plane.)

The dots in (\ref{modecom}) indicate terms that will not contribute to the 6-point function of $W$'s at order $c$. Composite currents are defined as
\be
\Lambda^{(4)} \sim LL \ , \quad \Lambda^{(5)} \sim LW \ , \quad \Lambda^{(6)} \sim WW \ \ , \quad \Theta^{(6)} \sim -L' W + \tfrac{2}{3}LW'\ ,
\ee
with the mode expansions
\begin{align}\label{nonlinearmodes}
\Lambda^{(4)}_{n} &= \sum :L_{n-p}L_p:\\
\Lambda^{(5)}_{n} &= \sum :L_{n-p}W_p:\notag\\
\Lambda^{(6)}_{n} &= \sum :W_{n-p}W_p:\notag\\
\Theta^{(6)}_n &= \sum (\tfrac{5}{3} p - n) : L_{n-p} W_p:\notag \ .
\end{align}
These composite currents are not quasiprimary, but they can be made so by adding 
linear corrections. Since these linear corrections involve fewer fields, their contribution
to the vevs will be subleading for large $c$, and hence do not affect our results.

\begin{singlespace}

\end{singlespace}

\end{document}